\documentclass[aip,jmp,bmf, sd, rsi, amsmath,amssymb, reprint]{revtex4-1}
\usepackage{graphicx}
\usepackage{bigints}

\begin{document}

\title{Theory of quasi-ballistic FET:
steady-state regime and low-frequency noise.}

\author{M. Yelisieiev}
\affiliation{Taras Shevchenko National University of Kyiv, Kyiv, Ukraine}
\email{mykola.eliseev@gmail.com}
\author{V. A. Kochelap}
\affiliation{Institute of Semiconductor Physics, NAS of Ukraine, Kyiv, Ukraine}

\begin{abstract}
We present the theoretical analysis of steady state regimes and
low-frequency noises in quasi-ballistic FETs.
The noise analysis is based on the Langevin approach, which accounts
for the microscopic sources of fluctuations originated from
intrachannel electron scattering. The general formulas for
local fluctuations of the carrier concentration, velocity and
electrostatic potential as well, as their distributions along the channel
are found as functions of applied voltage/current. Two circuit regimes
with stabilized current and stabilized voltage
are considered. The noise intensities for the devices
with different ballisticity are compared.

We suggest that the presented analysis makes better comprehension
of physics of electron transport and fluctuations in quasi-ballistic FETs,
improves their theoretical description and can be useful for device
simulation and design.
\end{abstract}

\maketitle

\section{Introduction}\label{Sec I}

In short channel field effect transistors (FETs), electrons
experience only a few collisions with defects and phonons during
the transient time, while for typical  electron concentrations electron-electron
collisions are dominating and cause hydrodynamic behavior of the electron gas.
For such a physical situation, Dyakonov and Shur have proposed
to model the electron transport as that of a charged fluid, which is confined in a narrow
layer and governed by  hydrodynamic equations.~\cite{D-Sh-1993}
The electrons are characterized by the area density, $n$, and
by the drift velocity, $v$, induced by  source-drain electric bias, $\phi$.
In the frame of the gradual channel approximation~\cite{gr-chan-appr-1,Mitin},
the local potential is supposed to be proportional to the electron density.

\begin{figure}
\includegraphics[scale=0.7]{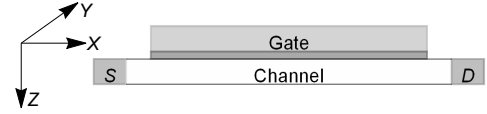}
\caption{A sketch of FET under consideration. }
\label{fig-1}
\end{figure}

The complete system of equations for the Dyakonov-Shur model of quasi-ballistic
FET reads:
\begin{gather}
\frac{\partial v}{\partial t} + v \frac{\partial v}{\partial x} +\frac{v}{\tau}
= \frac{e}{m} \frac{\partial \phi (x,h)}{d x}\,, \\
\frac{\partial n}{\partial t} + \frac{\partial j}{\partial x} = 0\,,\,\,\,j = n v\,,
\\
\phi(x,z) = -\frac{4 \pi e}{\kappa} n(x) z + \phi_g\,\,\,\,(0 \leq z \leq h).
\end{gather}
These equations are for the frame of reference presented in Fig.~\ref{fig-1},
where geometry parameters of the FET under consideration are indicated;
the conductive layer and the gate are situated at $z= h$ and $z=0$, respectively.
The voltage applied to the gate is $\phi_g$; $m$ and $-e$ are
the electron effective mass and the
electron charge, $\kappa$ is the dielectric constant, $\tau$ is an electron relaxation time,
$j $ is the electron flux density. Eq.~(3) obtained for the gradual channel
approximation is valid at $L_x \gg h$ and characteristic scale of $\phi(x)$ variation is
much larger than $h$.

In Dyakonov-Shur paper \cite{D-Sh-1993} and numerous subsequent
publications\cite{Cr-2000, Crown-1,Crown-2,Crown-3},
Eqs. (1) - (3) have been applied mainly for time-dependent problems focusing to
ultra-high frequency instabilities arising at specific boundary conditions
(i.e., at suitable microwave environment).

Meanwhile, for steady-state conditions,
 Eqs.~(1)-(3) permit the finding of {\it exact analytical solutions}.
 Analytical solutions independent on numerical methods are always important to make
 general statements, including qualitative conclusions.
 In this paper we obtain solutions for all variables
 $v(x),\,n(x),\,\phi(x)$,
present analysis of the current-voltage characteristics, particularly,
we clarify the origin of what is called "pinch-off" effect. On the base
of the steady-state solutions we studied
non-equilibrium electron fluctuations in the FET channel.

In general, analysis of current- and/or voltage noises in devices, including FETs,
is a very complex problem because of a number of factors affecting
electron fluctuations, among them highly nonuniform carrier distributions
and drift velocities along the active channels,
both are induced by applied voltages, etc.
These and other effects characteristic for nonlinear electron
transport in quasi-ballistic FETs are significant and improved theoretical analysis
is relevant from  viewpoints of the device physics and device
applications.

The model under consideration
and analytical solutions facilitate the noise analysis.
Below  we present the analytical study of
low-frequency noise in quasi-ballistic FETs.
The analysis is based on the Langevin approach, which accounts
for the microscopic sources of fluctuations originated from
intrachannel electron scattering. The general formulas for
local fluctuations of the carrier concentration, velocity and the
electrostatic potential, and their distributions along the channel
are found as functions of applied voltage/current.
This enabled to reveal the effect of electron correlations
under the metal gate on the fluctuations.
Two circuit regimes are considered: (A) stabilized
current regime (suppressed {\it ac} current fluctuations) and
(B) stabilized voltage regime (suppressed {\it ac} voltage fluctuations).
Respectively, the spectral densities of  voltage-noise and
current-noise in the FETs are derived.
Comparing results for the devices with different ballisticity degrees,
we have concluded, that at a given current, FETs with
larger ballisticity of the active channels demonstrates
larger low-frequency voltage noises for the circuit A and smaller
current noises for the circuit B relatively to more dissipative
channels.

\section{The steady-state solutions}

For the steady-state, Eq.~(2) gives for the electron flux density: $j= n\,v=j_0$ with $j_0$
being an integration constant. It is convenient to solve the system (1)-(3) at a
given $j_0$ and then to find the voltage drop on the device, $\phi(L_x)$. The boundary
condition for Eq.~(1) is $v(0) = j_0/n_s$, where $n_s$ is the electron area density
near the source ($n_s = n(0)$). For what follows, we introduce dimensionless
variables and parameters:
\begin{gather} \label{sc-1}
V=\frac{v}{v_s}\,,\,\,N:=\frac{n}{n_s}\,,\,\,\xi=\frac{x}{L_x}\,,\,\,\,\\
\Phi(\xi) = \frac{[\phi(\xi,-h)-\phi_g]}{u_{sc}}\,,\,\, \, u_{sc}=\frac{4 \pi e h n_s}{\kappa}\,, \label{sc-2} \\
J =\frac{j_0}{j_{sc}}\,,\,\,\, j_{sc} =\sqrt{\frac{4 \pi e^2 h n_s^3}{m \kappa}}\,,\,\,\,
{\cal B}= \frac{L_x}{\tau} \sqrt{\frac{m \kappa}{4 \pi e^2 h n_s}}\,. \label{J-B}
\end{gather}
Here the scaling parameters for the potential, $u_{sc}$, and the flux, $j_{sc}$, account for the
effect of interaction of the electrons with the metal gate. Factor $\cal B$ is the only parameter
dependent on kinetic characteristic, which is the electron relaxation time, $\tau$.

In these notations we obtain the equation for $N$:
\begin{equation}  \label{eq-N}
\frac{N^3-J^2}{N^2}\,\frac{d N}{d \xi} = - J {\cal B}\,
\end{equation}
with $N(0)=1$. Solution of Eq.~(\ref{eq-N}) gives implicit dependence $N(\xi)$:
\begin{equation}  \label{N-x}
J^2 \left(1-\frac{1}{N}  \right) + \frac{1- N^2}{2} = J {\cal B}\,\xi\,.
\end{equation}
Setting $\xi=1$ in the latter equation, we find the dimensionless concentration
at the drain, $N (1)$, and the voltage drop on the conductive channel,
$U=\Phi(1) - \Phi (0)=N(0)-N(1) \equiv 1-N(1)$. Then, Eq.~(\ref{N-x}) leads to the following
relationship between $J $ and $U$:
\begin{equation}\label{I-V-1}
{\cal L}(J,U) \equiv \frac{U(2-U)}{2 J}- J \frac{U}{1-U}={\cal B}\,.
\end{equation}
At a given $J$, the function ${\cal L}(J,V)$ has maximum at $U=1 -J^{2/3}$. Thus,
the following equation
\begin{equation} \label{Jc}
max{\cal L} = J_c+\frac{1}{2 J_c} -\frac{3}{2} J_c^{1/3} = {\cal B}
\end{equation}
determines maximal possible $J=J_c ({\cal B})$ and corresponding voltage drop
$U_c=1-J_c^{2/3}$ allowable in the model under consideration.
Eq.~(\ref{I-V-1}) has solutions (two solutions) only at $J \le J_c (B)$.
The case $ {\cal B}\ll 1$ corresponds to almost ballistic electron
transport. For them $J_c \approx 1- \sqrt{3 {\cal B}/2}$. The
opposite case, $ {\cal B}\gg 1$, is relevant to dissipative transport with
$J_c \approx 1/2\,B$. Yet, for parameters allowing the solutions there exists
additional restriction: $J  {\cal B}\leq 1$.
Of two branches of the dependence $J(U)$ we shall select that corresponding to
a positive flux ($J,V>0$). This gives the dimensionless current-voltage
characteristic:
\begin{equation} \label{I-V-2}
J = -\frac{ {\cal B}(1-U)}{2 U} + \sqrt{\frac{{\cal B}^2 (1-U)^2}{4 U^2}
+ \frac{(1-U)(2-U)}{2}}\,.
\end{equation}
\begin{figure}
\includegraphics[scale=0.5]{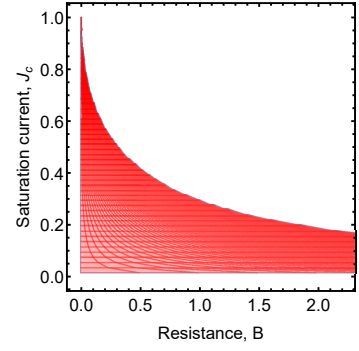}
\caption{Dimensionless saturation current $J_c$ as a function of parameter
${\cal B}$}
\label{fig-2}
\end{figure}

\begin{figure}
\includegraphics[scale=0.3]{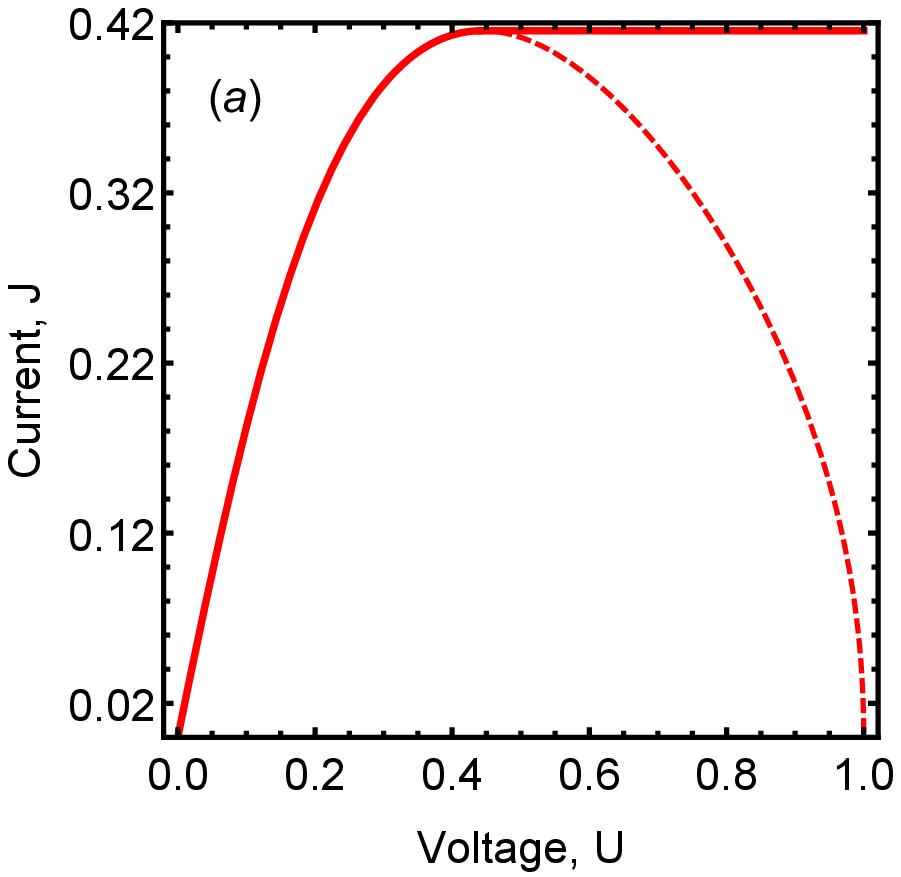}
\includegraphics[scale=0.3]{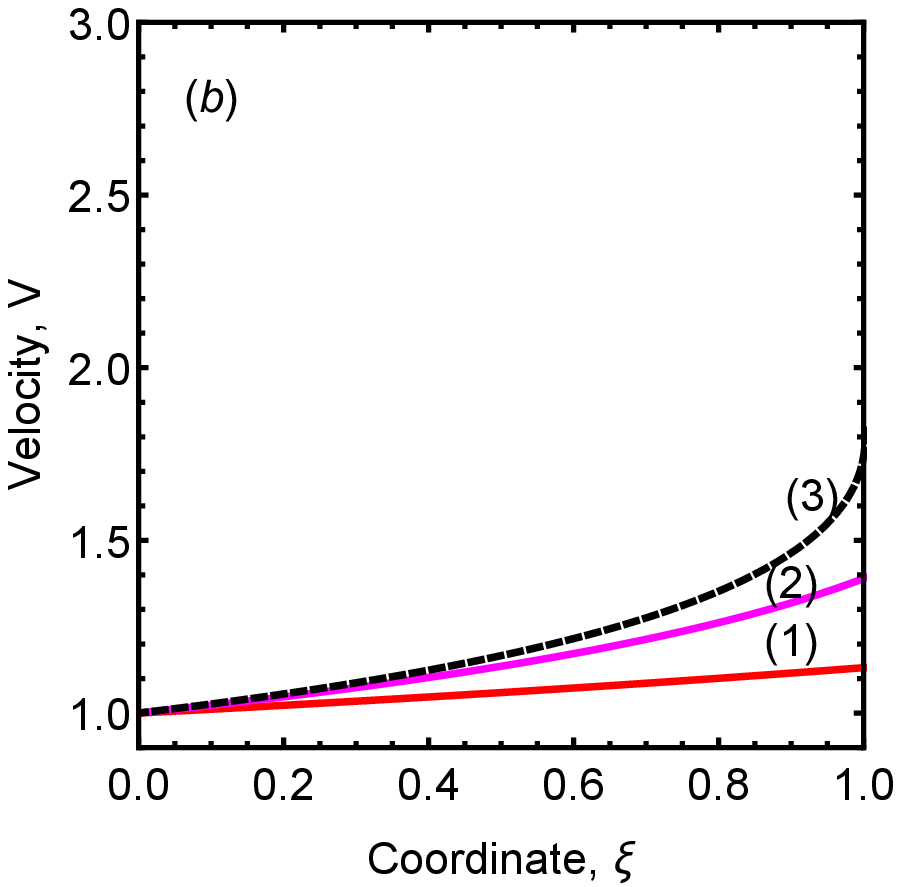}
\includegraphics[scale=0.3]{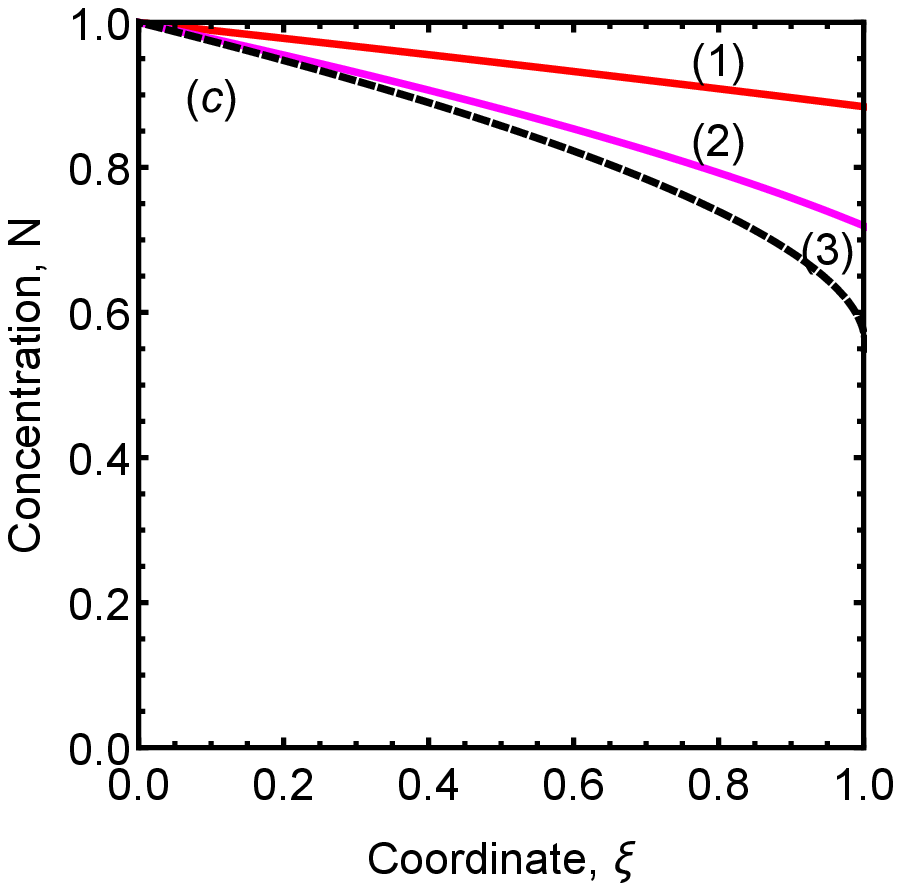}
\includegraphics[scale=0.3]{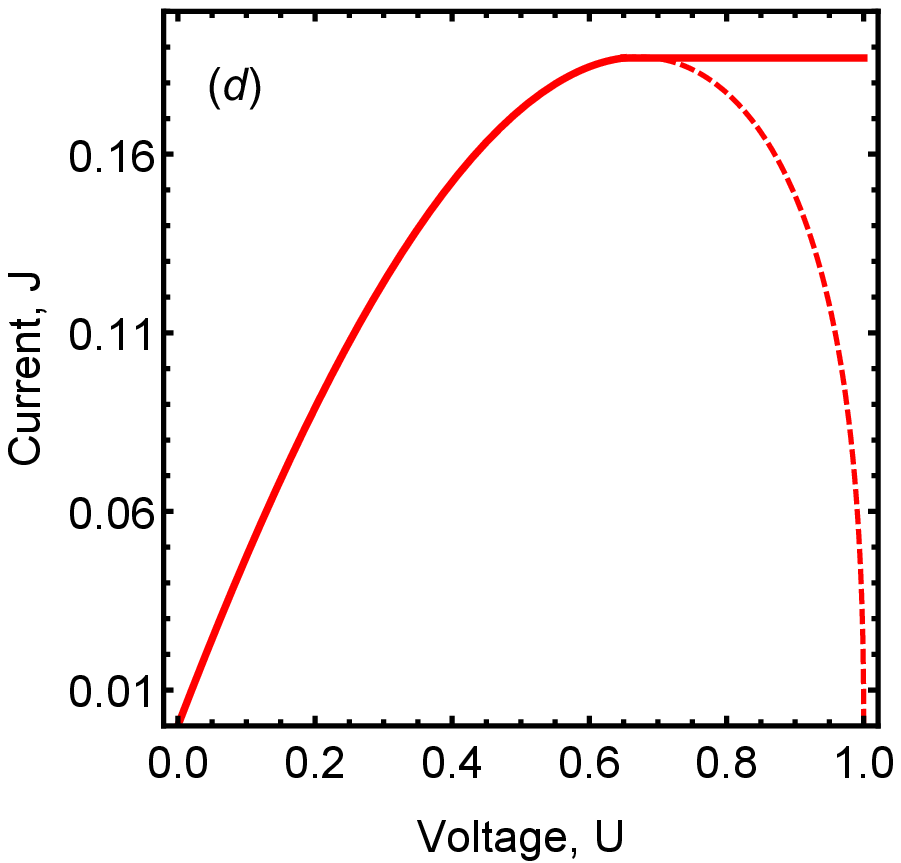}
\includegraphics[scale=0.3]{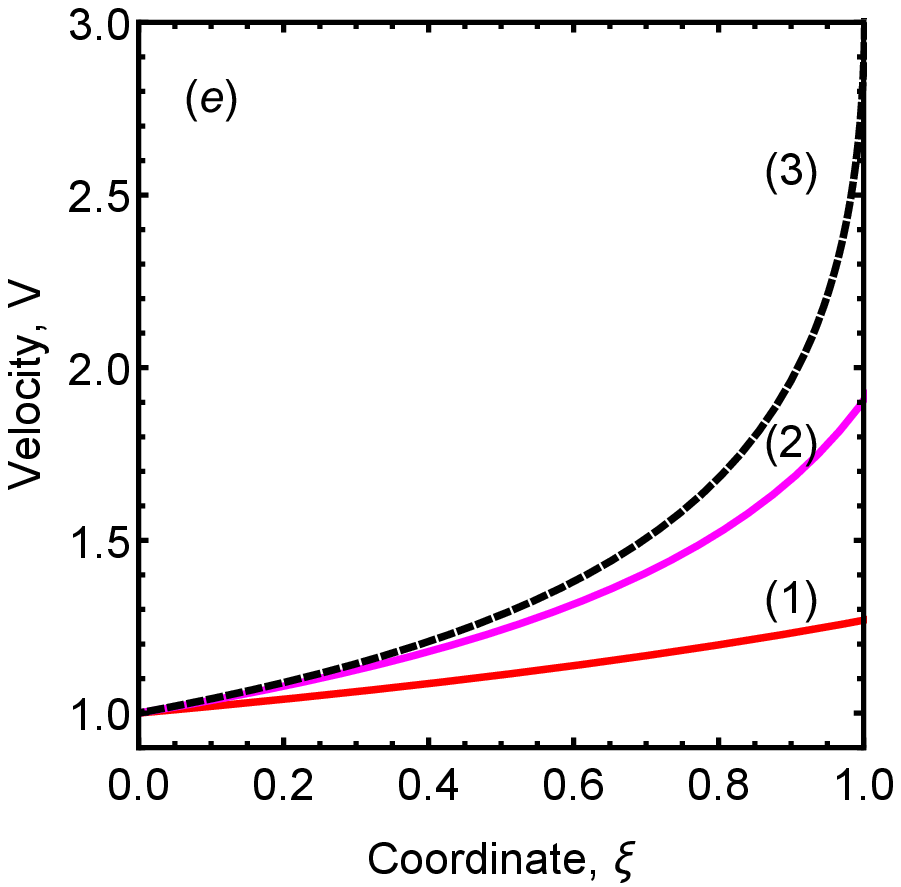}
\includegraphics[scale=0.3]{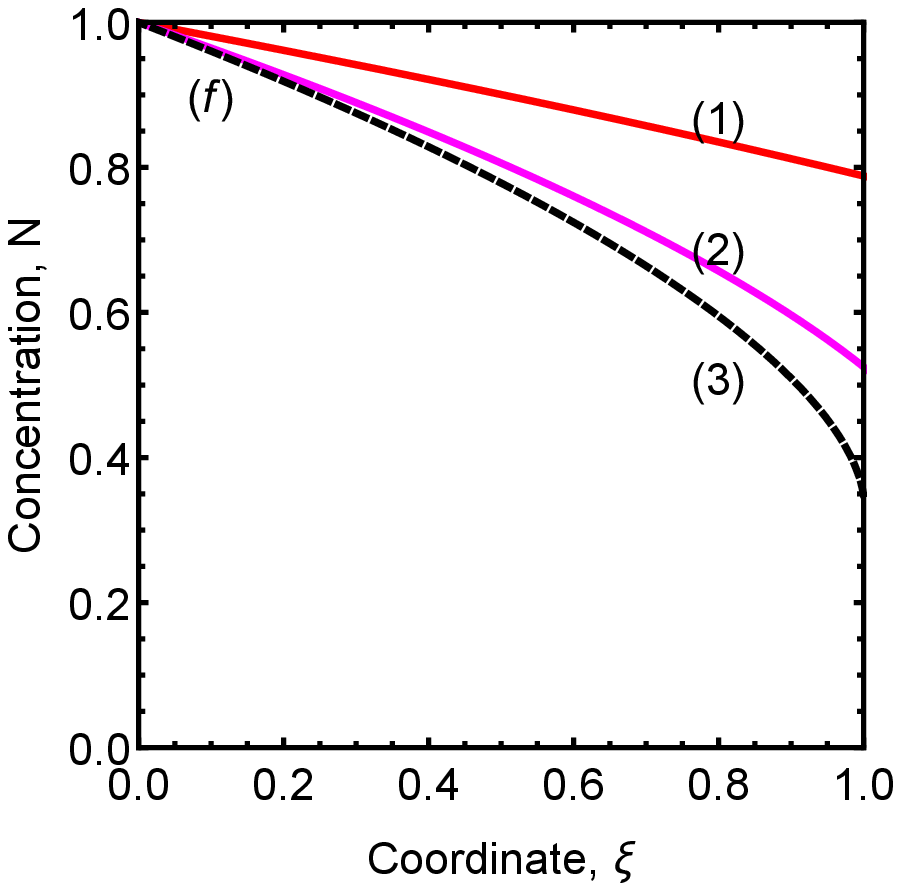}
\caption{Dimensionless characteristics of quasi-ballistic FETs
for ${\cal B} =0.5,\,J_c=0.415$, panels (a), (b), (c) and
${\cal B} =2,,Jc=0.187$, panels (d). (e), (f).
(a), (d): Current-voltage characteristics; current saturation portions are shown conditionally.
(b), (e): Dimensionless velocities, $V (\xi)$. (c), (f): Electron concentrations, $N(\xi)$.
Curves $1,\,2$ and $3$ correspond to currents $0.5\,J_c,\,0.9\,J_c$ and $J_c$. Curves 3 on panels (c), (f) clearly demonstrate
the absence of real pinch-off effect in quasi-ballistic FETs.}
\label{fig-3}
\end{figure}

In Fig.~\ref{fig-3} we present the characteristics of the FETs
for two values of $\cal B$ (${\cal B}=0.5\,$ and ${\cal B}=2$).
The current-voltage characteristics
allowed in the model are shown in the panels (a) for $J \leq J_c(0.5)= 0.41$,
$U \leq U_c(0.5) =0.43$ and $J \leq J_c(2)= 0.187$, $U \leq U_c(2) =0.67$.
At small voltage bias, Eq.~(\ref{I-V-2}) gives the following current dependence:
\begin{equation} \label{I-V-3}
J \approx \frac{U}{{\cal B}} - \frac{U^2}{2 {\cal B}} -\frac{U^3}{{\cal B}^3} ...\,.
\end{equation}
Thus, parameter ${\cal B}$ is the dimensionless FET resistance in the linear regime.
At arbitrary $J$
and $U$ the {\it dimensionless} differential resistance can be found in the
parametrical form:
\begin{equation} \label{d-resist}
{\cal R}_D =
\frac{N(1) [1-N(1)] [2 J^2-N(1) (1+N(1))]}{2 J[J^2-N^2(1)]},
\end{equation}
where $N(1)= 1 - U$ and $J(U)$ is given by Eq.~(\ref{I-V-2}).
At large $U$, i.e., beyond the model applicability,
the currents should saturate.
In Fig.~\ref{fig-3}(a)  the saturation portions are shown
conditionally. Note, from Fig.~\ref{fig-2} it follows that the critical
current $J_c$ decreases with increasing of parameter
${\cal B}$ (i.e., at larger $L$ and/or smaller $\tau$).

In the panel (b) and (c) of these figures,
distributions of the electron velocities and densities along the channel  are
presented. These distributions are linear near the source:
\begin{equation} \label{N-V-source}
N(\xi) \approx 1- \frac{J {\cal  B}}{1-J^2}\, \xi\,,\,\,\,
V(\xi) = 1 + \frac{J {\cal B}}{1-J^2}\,\xi\,.
\end{equation}
 At larger $\xi$ the distributions, generally, are nonlinear (concave and convex
 dependencies, respectively). As $J \rightarrow J_c$, one can find:
 \begin{gather} \nonumber
 N (\xi) \approx J_c^{2/3} + \frac{2 }{3 J_c^{1/3}} {\cal B}^{1/2} \sqrt{1- \xi}\,, \\
 \nonumber
 V (\xi) \approx J_c^{1/3} - \frac{2 }{3 J^{2/3}} {\cal B}^{1/2} \sqrt{1- \xi}\,.
 \end{gather}
 Similarly, the potential distributions near the source and the drain are
 \begin{gather} \nonumber
 \Phi(\xi) \approx \Phi (0) + \frac{J {\cal B}}{1-J^2}\, \xi\,,\,\,\,(\xi \ll 1)\,\,\\
 \Phi(\xi) \approx  J_c^{2/3} + \frac{2}{3 J_c^{1/3}} {\cal B}^{1/2}
  \sqrt{1- \xi}\,,\,\,\,\,
 (\xi \rightarrow 1,\,\,J \rightarrow J_c).
 \nonumber
 \end{gather}
Therefore, we find that at
the critical current, $J_c$, the electron density is nonzero everywhere
 in the conductive channel,   and the velocity remains a finite  value.
The former means that in quasi-ballistic FETs actually there is no pinching-off
of the conductive channel. The pinch-off effect ($N (1) \rightarrow 0$) evidences
at large ${\cal B}$ (long channel, $L \rightarrow \infty$,
 and/or strongly dissipative transport,
  $\tau \rightarrow 0$). While the electric field in the channel
diverges in this limit: $\left| \frac{d \Phi}{d \xi }\right| \rightarrow \infty$
at $J \rightarrow J_c$.

\section{Low-frequency noise}

Among different sources generating current/voltage noises in FETs,
we focus on the low-frequency noise of an intrinsic nature, namely, that caused
by random character of electron scattering in the conductive channel of FET.
In the frame of the Langevin approach~\cite{Van-Fleet,Kogan,Landau-St-M}
this type of noise can be evaluated by the use of linearized Eq.~(1) supplemented
with a random force $f(x,t)$. The ensemble average
of the latter function is $\overline{f(x,t)} =0$ and its properties are determined
by a correlator $ \overline{f(x,t)f(x',t')}$, for which it is assumed that~\cite{Kogan}
\begin{equation} \label{corr-1}
 \overline{f(x,t)f(x',t')} = g(x, t)\, \delta(x-x')\, \delta(t-t')
\end{equation}
with $g(x,t)$ defined by random scattering and a local density,
and velocity distribution of the carriers. Here we consider the fluctuations around
steady-state of the FET,
thus dependence on time is absent: $g(x, t) = g(x)$.
Introducing $f_{\omega} (x)$ as Fourier transformation
of the force $f(x,t)$, one can transform Eq.~(\ref{corr-1}) to the form:
\begin{equation}  \label{corr-2}
\overline{f_\omega (x) f_{\omega'} (x)} = \frac{1}{2 \pi} \,g (x)\, \delta(x-x') \,\delta(\omega+\omega')\,.
\end{equation}

We present the variables as
$v+v_1,\,n+n_1,\,\phi+\phi_1$, with $v_{1},\,
n_{1},\,\phi_1$ being the fluctuation values.
The Langevin equation reads
\begin{equation} \label{L-1}
\frac{\partial v_1}{\partial t} + v_1 \frac{\partial v}{\partial x}
+ v \frac{\partial v_1}{\partial x} +\frac{v_1}{\tau} = \frac{e}{m} \frac{\partial
\phi_1 (x,h)}{d x} + \frac{1}{m} f(x,t)\,.
\end{equation}
Two other equations for $n_1$ and $\phi_1$ can be obtained from Eqs.~(2), (3)
substituting $n,\,\phi \rightarrow n_1,\,\phi_1$ and $j \rightarrow j_1=n v_1+ n_1
v $. It is convenient to introduce the dimensionless fluctuation values
$N_1(\xi,t),\,V_1(\xi,t),\,\Phi_1(\xi,t)$, as done in relationships (4), (5).
For the noise characteristics we define the Fourier transformations for all variables:
$N_\omega(\xi),\,V_\omega(\xi),\,\Phi_\omega(\xi),\,J_\omega,\,U_\omega$ and $f_\omega(\xi)$.
Generally, temporal current/voltage fluctuations are dependent on microwave
environment of the device. However the master equation for the fluctuations can be
found for arbitrary  external electric circuits, which provides $J_\omega \neq 0$ and $U_\omega \neq 0$.
Then, the fluctuation flux normalized to $j_{sc}$ is
\begin{equation}
J_\omega = \frac{j_1}{j_{sc}} =V_\omega N + N_\omega V \,.
\end{equation} \label{J-omega-N}
For low-frequency fluctuations one can drop derivatives with respect to
time, as well as terms proportional to $\omega$ in  equations for the Fourier
components. This is valid for $\omega \tau \ll 1$ and $\omega \ll v_s/L_x$.
As the result, we obtain the following equation in terms of $V_\omega$:
\begin{gather} \label{V-omega} \nonumber
\frac{d V_\omega}{d \xi} - J  {\cal B}\frac{3 N^4}{\left[J^2-N^3 \right]^2}\,V_\omega =\\
\frac{\tau }{m v_s} \frac{J  {\cal B} N}{J^2 - N^3} f_\omega(\xi) -  \frac{J  {\cal B} N^3}{[J^2-N^3]^2} J_{\omega}  \,. \label{V-omega}
\end{gather}
This is nonhomogeneous differential equation of the
first order with $\xi$-dependent coefficients expressed through the steady-state
solution $N(\xi)$:
\begin{figure}
\includegraphics[scale=0.6]{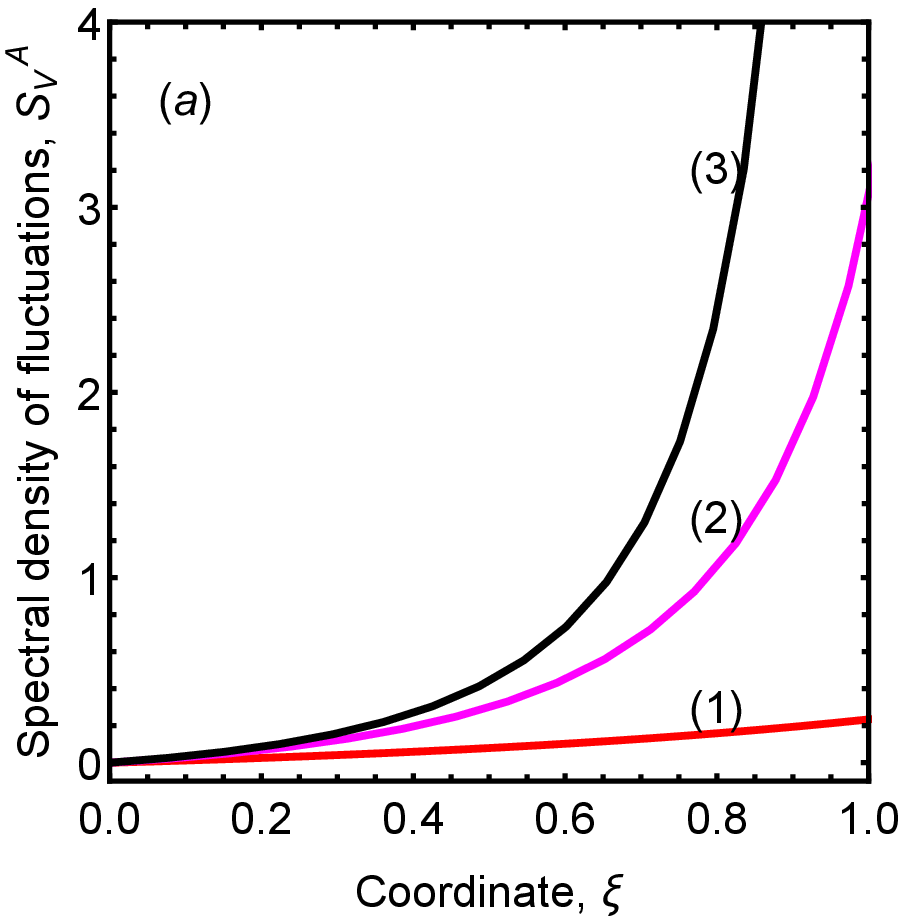}
\includegraphics[scale=0.6]{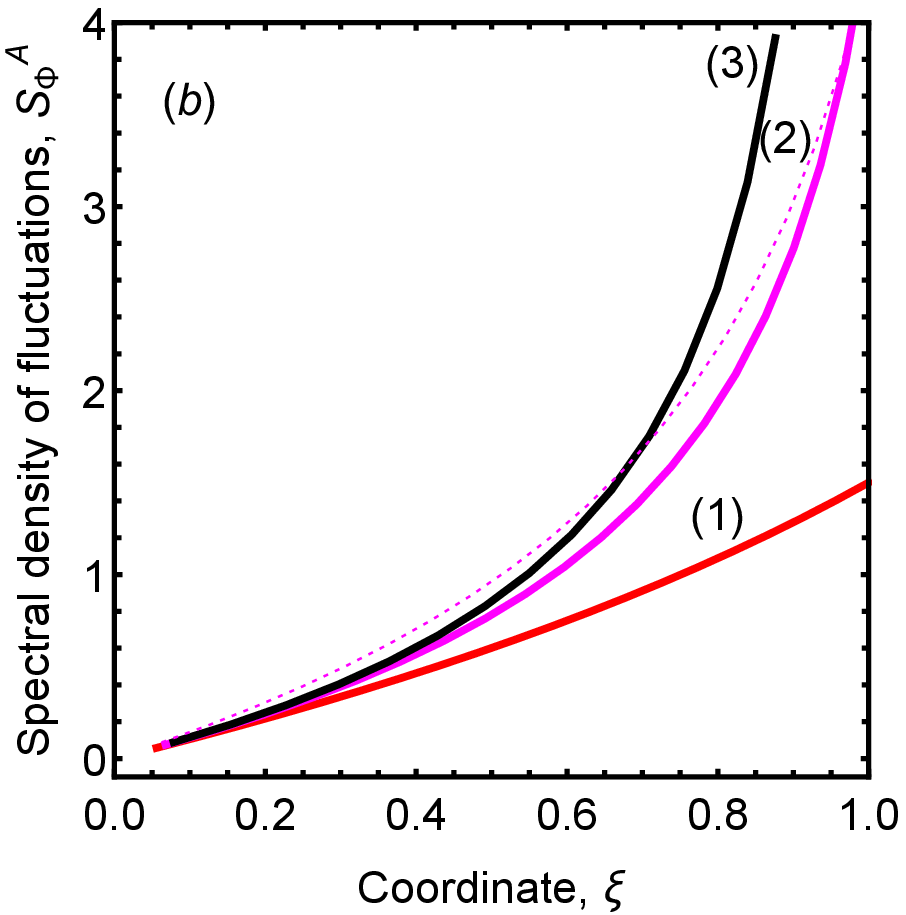}
\includegraphics[scale=0.6]{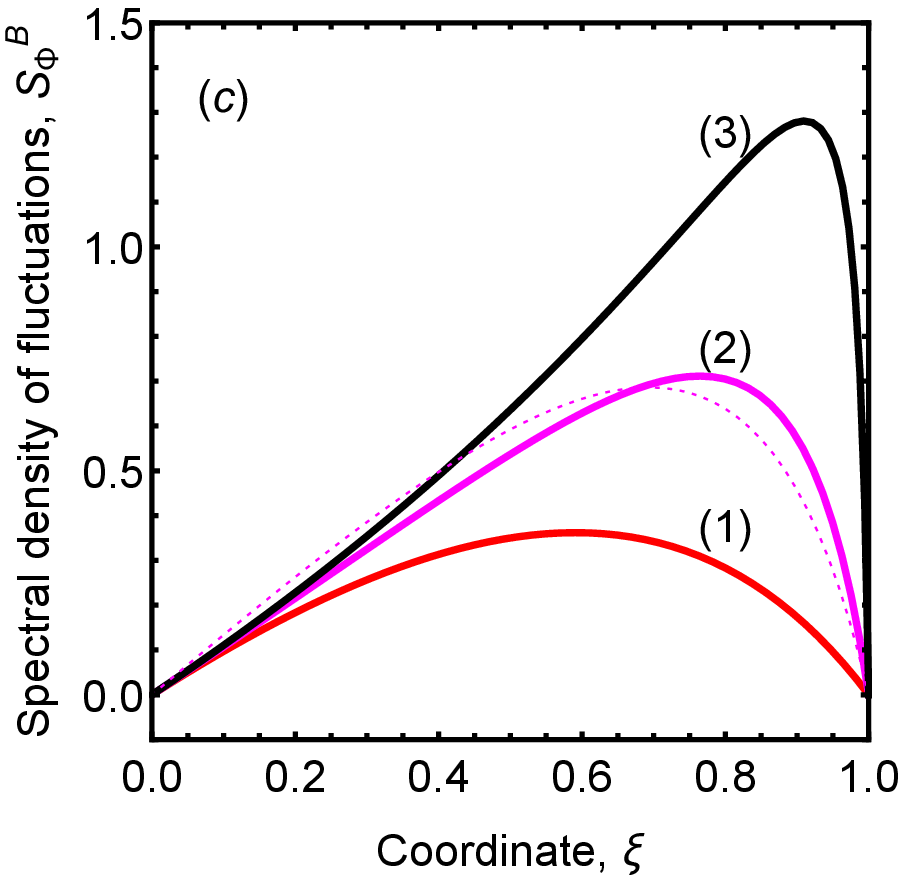}
\caption{Spatial distributions of spectral densities of fluctuations
in FET for ${\cal B}=0.5$; (a): $s^{A}_{V_{\omega}} (\xi)$;
(b) and (c): $s^A_{\phi_{\omega}} (\xi)$ and $s^B_{\phi_{\omega}} (\xi)$
for the circuits A and B, respectively.
Curves $1-3$ are for corresponding currents indicated in Fig.~\ref{fig-3}.
In (b), (c), dotted lines are given for comparison at ${\cal B}=2$ and $J=0.9*Jc$.
Presented spectral densities $s_{\phi_{\omega}}$ are scaled to
the relevant Nyquist quantities given by Eq.~(\ref{phi-1}).
Notice, amplitudes of $s_{\phi_{\omega}}$ in panels (b) and (c) are very different.}
 \label{fig3B}
\end{figure}
\begin{gather} \label{V-omega-2}
\frac{d V_\omega}{d \xi} - P(\xi) V_\omega = Q(\xi) f_{\omega}+ W(\xi) J_{\omega}\,,
\end{gather}
with
\begin{gather} \nonumber
P(\xi) =  J  {\cal B}\frac{3 N^4(\xi)}{\left[J^2-N^3(\xi) \right]^2}\,,\\
Q(\xi)= \frac{\tau n_s }{m j_0} \frac{J  {\cal B}N(\xi)}{J^2 - N^3(\xi)} \,,\,\,\,
W(\xi)=- \frac{J  {\cal B}N^3(\xi)}{[J^2-N^3(\xi)]^2}\,. \nonumber
\end{gather}
Restricting ourselves by the intrinsic sources of the
fluctuations, we set the boundary condition to this equation in the form:
$V_\omega(0)=0$ (short-circuited gate-drain part of the circuit at finite
frequencies), for Eq.~( \ref{V-omega-2})
this tells $V_{\omega} (0) = J_{\omega}/ N(0)=J_{\omega}$.
Now, we easily find the solution of Eq.~(\ref{V-omega-2}):
\begin{gather} \nonumber
V_\omega(\xi) = J_{\omega} e^{\int_0^{\xi} d \xi' P(\xi')}+\\
 \int_0^{\xi} d \zeta\, \left[ Q (\zeta)f_{\omega}(\zeta) + W(\xi) J_{\omega} \right]
e^{-\int^{\zeta}_{\xi} d \xi' P(\xi')}\,.  \label{sol-1}
\end{gather}
The fluctuation flux, $J_{\omega}$,
can be determined, when the external electric circuit is specified.

\subsection{The circuit with suppressed current fluctuations
(the  stabilized current regime).}

For this case we set $J_{\omega}=0$
and Eq.~(\ref{sol-1}) can be rewritten in the form:
\begin{gather}
V^A_{\omega} (\xi)=
 \int_0^{\xi} d \zeta\, Q (\zeta) f_{\omega} (\zeta)
e^{-\int^{\zeta}_{\xi} d \xi' P(\xi')}  \equiv \nonumber \\
\int_0^{\xi} d \zeta \,{\cal K}(\xi,\zeta) f_{\omega} (\zeta)\,.  \label{sol-V}
\end{gather}
(The results for fluctuations obtained in this Subsection are labeled by
the upper mark $'A'$.)
Using dependence $N(\xi)$ implicitly given by Eq.~(\ref{N-x}) we can calculate
the integrals in Eq.~(\ref{sol-1}) and  the value ${\cal K}(\xi,\zeta)$
in terms of $N(\xi),\,N(\zeta)$. Indeed,  according to Eq.~(\ref{eq-N})
 $$
 d \xi = \frac{J^2-N^3}{J  {\cal B}N^2} d N\,
 $$
 we can change integration over $\xi'$ by that over $N$:
 $$
 \int_{\zeta}^\xi P(\xi')\,d \xi'  =\int_{N(\zeta)}^{N(\xi)} \frac{3 N^2\, d N}{J^2- N^3} =
 ln\left[\frac{J^2-N^3(\zeta)}{J^2-N^3(\xi)}\right]\,
  $$
  and find
  \begin{equation}
  {\cal K} (\xi,\zeta) =  \frac{\tau n_s J {\cal B}}{m j_0}
  \frac{N(\zeta)}{\left(J^2 - N^3(\xi)\right)} \equiv {\cal K}_N[N(\xi), N(\zeta)]\,.
  \end{equation}
The correlator for the velocity fluctuations is
$$
\overline{V^A_\omega(\xi)V^A_{\omega'}(\xi')} = \int_0^{\xi} \int_0^{\xi'}
d \zeta d \zeta' {\cal K} (\xi,\zeta){\cal K}(\xi', \zeta')
\overline{f_\omega(\zeta) f_{\omega'}( \zeta')}\,.
$$
Here $\overline{f_\omega(\zeta) f_{\omega'}( \zeta')} $ can be obtained from Eqs.~(\ref{corr-2}) and (\ref{A-4}) by substitution
$x,\,x' \rightarrow \zeta L_x, \zeta' L_x $:
$$
\overline{f_\omega(\zeta) f_{\omega'}( \zeta')} = \frac{1}{2 \pi} \overline{g} (\zeta) \delta(\zeta-\zeta') \delta(\omega + \omega')\,
$$
with
$$
\overline{g} (\zeta)\equiv \frac{1}{L_x} g(\zeta L_x) = \frac{2 D m^2}{\tau^2 n(\zeta) L_x L_y}\,.
$$
The above obtained correlator can be rewritten as
\begin{equation} \label{corr-3}
\overline{V^A_\omega(\xi)V^A_{\omega'}(\xi')} = \frac{1}{2 \pi} \delta(\omega+\omega')
 \int_0^{\xi} d \zeta \, \overline{g} (\zeta)\,{\cal K} (\xi,\zeta){\cal K}(\xi', \zeta) \,.
\end{equation}
Using Eqs.~(\ref{W-K-1}), (\ref{W-K-2}) we obtain the spectral
density of the electron velocity fluctuations in a point $\xi$ in the form:
\begin{gather} \nonumber
S^A_{V_{\omega}}(\xi) = 2
 \int_0^{\xi} d \zeta \, \overline{g} (\zeta)\,[{\cal K} (\xi,\zeta)]^2 =\,\\
 \frac{2}{J  {\cal B}} \int_1^{N(\xi)} dN \frac{J^2-N^3}{N^2}
 \overline{g}  [N]\,\left({\cal K}_N [N(\xi),N] \right)^2 \,. \label{S_V}
\end{gather}
Here $\overline{g} (\zeta)$ is expressed through $N(\zeta)$ using implicit dependence of Eq.~(\ref{N-x}): $ \overline{g}  [N]
\equiv \overline{g} (\zeta(N))$.
Calculation of the integral in Eq.~(\ref{S_V}) gives
\begin{gather}   \nonumber
S^A_{V_{\omega}}(\xi) = \frac{4 D\, n_s }{ j_0^2 L_x L_y} \,\Psi^A(\xi,J)\,,\,\,\\
\Psi^A (\xi,J) \equiv J  {\cal B} \,
\frac{1- N^3(\xi) +3 J^2 lnN(\xi)}{3\, [J^2 - N^3(\xi)]^2}\,. \label{S_V2}
\end{gather}
The spatial distribution of the spectral density of the fluctuations of the  dimensional velocity, $v_{\omega}$,  is given by
\begin{equation} \label{S_v}
s^A_{v_{\omega}}(\xi) = \frac{4 D}{ n_s L_x L_y} \Psi^A (\xi,J)\,.
 \end{equation}
 At small $J$ (or $U$), we find this spatial distribution in the simple form
 \begin{gather} \nonumber
 s^A_{v_{\omega}}(\xi) \approx  \frac{4 D (JB)^2}{n_s L_x L_y}  (\xi +15 J {\cal B}\xi^2 +... ) = \\
 \frac{4 D\, U^2}{n_s L_x L_y}  \left[\xi + \left(\frac{1}{2} \xi+ \xi^2 \right) U + ... \right] \,.  \nonumber
 \end{gather}
 That is, the spectral density of the velocity fluctuations increases along the electron flux
 and the rate of this increase is quadratic on $J$ (or $U)$ at small currents.
 For finite currents, the spatial distributions
 of the velocity fluctuations are presented in Fig.~\ref{fig3B} (a). The fluctuations increase considerably at
 the drain side of the conductive channel. At $J \rightarrow J_c$ and $\xi \rightarrow 1$, Eq.~(\ref{S_v}) predicts infinitely
 high fluctuation intensity:
$$
s^A_{v_{\omega}} (\xi) \propto \frac{1}{\left[J_c - J - 2\,
\sqrt{{\cal B}\,(1-\xi)}\right]^{2}}\,.$$

Eq. (\ref{S_V}) facilitates finding the spectral density
fluctuations of the concentration and the potential
as functions of $\xi$: $S_{N_{\omega}} (\xi) =S_{\Phi_{\omega}} (\xi)
= N^4(\xi) S_{V_{\omega}} (\xi)$. These dependencies
are presented in Fig.~(\ref{fig3B}) (b).
In the dimensional form for the spectral density fluctuations of the potential we obtain
\begin{gather} \nonumber
s^A_{\phi_{\omega}} (\xi)= \left(\frac{4 \pi e h n_s}{\kappa} \right)^2 S_{\Phi_{\omega}} =
s_u^{(Nyq)}\frac{N^4(\xi) }{(J B)^2} \Psi^A (\xi,J)\,,\\
s_u^{(Nyq)} =\frac{4 D  m^2 L_x}{e^2 n_s \tau^2 L_y}\,. \label{phi-1}
\end{gather}
Bellow we will show that the quantity $s_u^{(Nyq)}$ coincides with the Nyquist
density of voltage-noise for the active channel.
At small $J$ (or $U$) we find
$$
s^A_{\phi_{\omega}} (\xi) \approx s_u^{(Nyq)} (\xi + J  {\cal B}\xi^2) =
s_u^{(Nyq)}(\xi + U \xi^2)\,.
$$
For finite currents the spatial distribution of the potential fluctuations in
the channel is similar to that of previously studied spatial distributions of
the velocity fluctuations, including infinite growth of these fluctuations at
$J \rightarrow J_c,\,\,\xi \rightarrow 1$.

Fluctuations of the total voltage drop in the FET, $u_{\omega} =\equiv \phi_{\omega}(1)$, are of
a finite value at  small currents:
\begin{equation} \label{phi-2}
s^A_{u_{\omega}} =s_u^{(Nyq)}
\left(1 + \frac{3}{2}\, U + ... \right) \, \,\,\,\,(U,\,\,J \rightarrow 0)\,,
\end{equation}
they increase with the current/voltage and become infinitely high at $J \rightarrow J_c$,
as illustrated in Fig.~\ref{fig-5} (a).

\subsection{The circuit with suppressed voltage fluctuations
(the voltage stabilized regime).}

For this case the fluctuations of the total voltage drop is zero,
$U_{\omega}=\Phi_{\omega}(1) = -N_{\omega}(1) =0$. This leads to
the condition $V_{\omega} (1) =J_{\omega}/N(1)$, which can be satistied at
\begin{gather}
J_{\omega} = \frac{N(1)}{\Delta}{\int_0^1 d \zeta Q (\zeta) f_{\omega} (\zeta)
e^{-\int^{\zeta}_{1} d \zeta' P(\zeta')} }= \frac{N(1)}{\Delta} \,V^A_{\omega} (1)\,,
\nonumber \\
\Delta =
{1- N(1) \left[ e^{\int_0^1 d \xi P(\xi)}
 + \int_0^1 d \zeta W(\zeta) e^{-\int^{\zeta}_1 d \zeta' P(\zeta')} \right]}
 \,, \nonumber
 \end{gather}
 where $V_{\omega}^A (1)$ is given by Eq.~(\ref{sol-V}) for $\xi=1$.
The above relationships allow easy
to find the correlator  $\overline{J_{\omega} J_{\omega'}}$, and
the spectral density of the total current fluctuations in the
FET for the circuit with the voltage stabilized regime:
\begin{equation} \label{S-J}
S^B_{J_{\omega}} = \frac{N^2(1)}{\Delta^2} \, S^A_{V_{\omega}}(1)\,
\end{equation}
with $S^A_{V_{\omega}}(\xi)$ defined by Eq.~(\ref{S_V2}).
(The results obtained in this Subsection are labeled by the upper mark $'B'$.)
Calculations give us the spectral density of the current fluctuations in
the analytical form
\begin{gather} \label{s-j}
S^B_{J_{\omega}} =  \frac{4 D n_s}{j_0^2 L_x L_y} \Psi^B (J)\,,\\
\Psi^B (J)\equiv \frac{4}{3} J  {\cal B}\frac{N^2(1) \left[1-N^3(1)+ 3 J^2 ln(N(1) \right]}
{[1-N(1)]^2 [2 J^2 + N(1) (1 + N(1))]^2}\,. \nonumber
\end{gather}
Together with relationship $N(1)=1-U$ and Eq.~(\ref{I-V-2}), these give the
voltage/current dependence of the current fluctuations for the considered case.
At small $U$ (or $J$) Eq.~(\ref{S-J}) gives
\begin{equation}  \label{S-J-0}
S^B_{J_{\omega}} = \frac{4 D n_s}{j_0^2 L_x L_y} \left(1-\frac{1}{2}\, U \right)\,
\,\,\,(J,\,U \rightarrow 0).
\end{equation}

\begin{figure}
\includegraphics[scale=0.6]{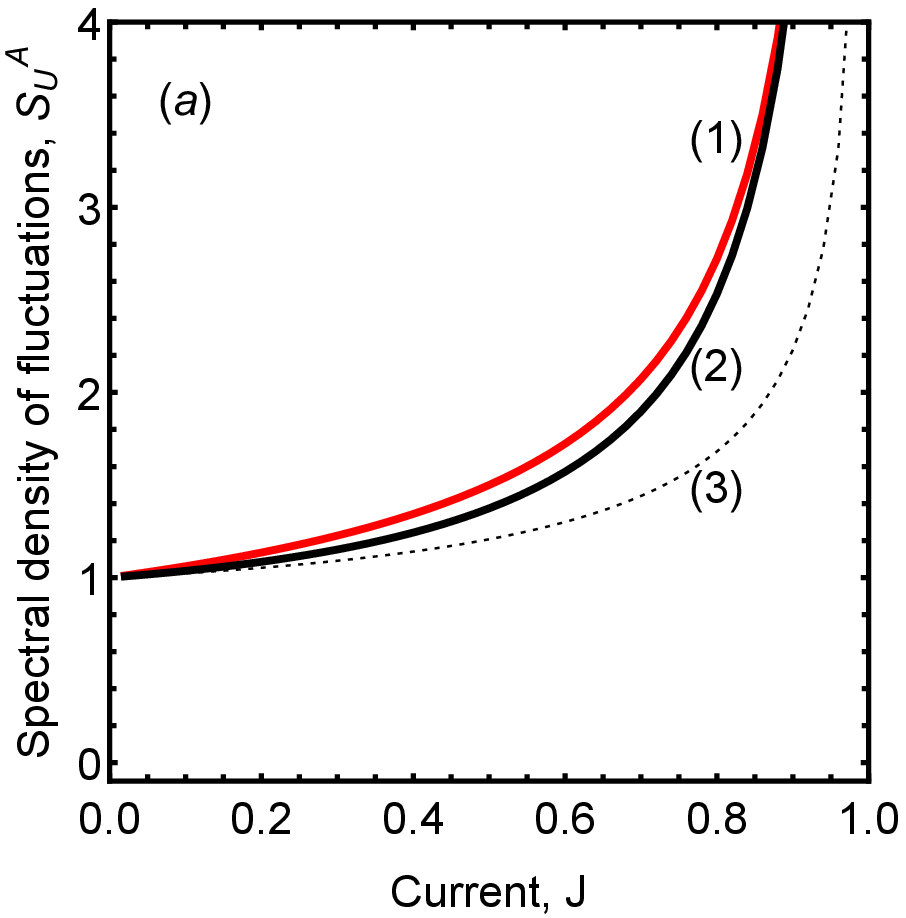}
\includegraphics[scale=0.6]{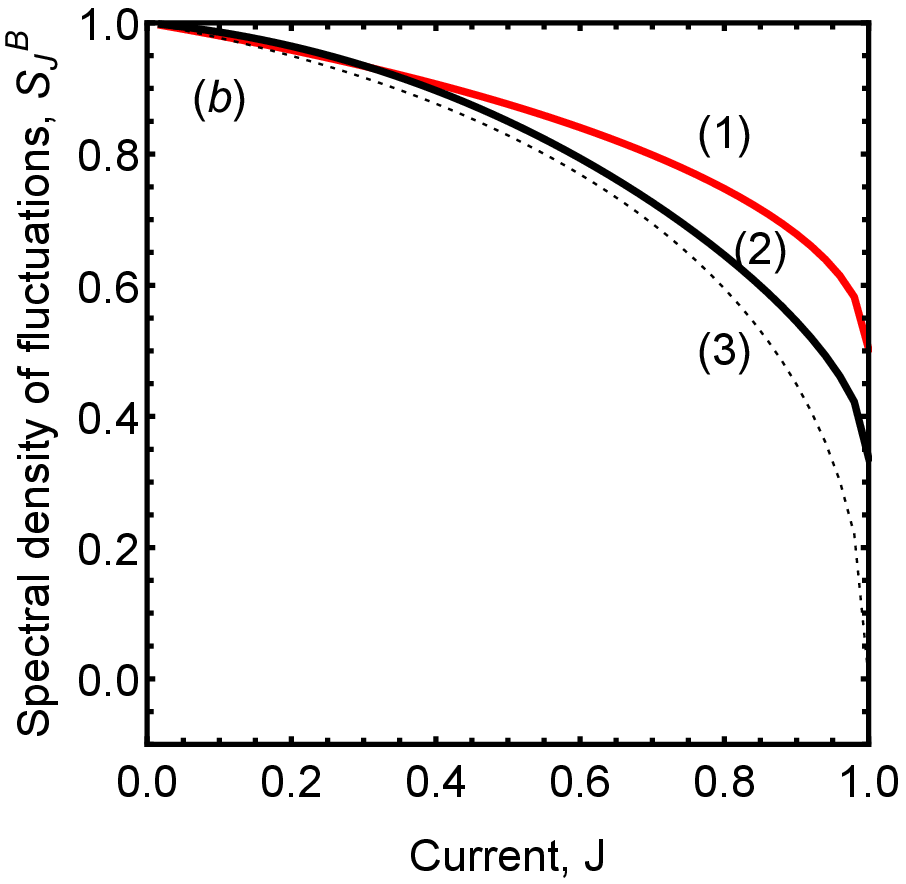}
\caption{The spectral densities of voltage and current fluctuations in FET  (full lines).
(a):  $s^A_{u_{\omega}}$ for the circuit A; (b):  $s^B_{J_{\omega}}$
for the circuit B; curves 1, 2 are for ${\cal B} =0.5,\,2$, respectively.
The spectral densities are scaled to corresponding Nyquist quantities (\ref{phi-1}), (\ref{s-j-3})
and shown as functions of $J/J_{c}$. Dotted lines (3) are relevant estimates with
the use of the differential resistance of Eq.~(\ref{d-resist}).}
\label{fig-5}
\end{figure}

For finite value of the applied current/voltage,  the fluctuation spectral densities
are illustrated by Figs.~\ref{fig-5} (b).
The dimensional form of the spectral density of the current
fluctuations in the active channel of the width $L_y$ can be recovered as follows:
\begin{equation} \label{s-j-2}
s^B_{j_{\omega}} =e^2 j_0^2 L_y^2 S^B_{J_{\omega}}\,.
\end{equation}
In particular, from Eq.~(\ref{S-J-0}) it follows, that at small currents/voltages
the spectral density of current fluctuations is equal
\begin{equation} \label{s-j-3}
s_j^{(Nyq)} = 4 D e^2 n_s \frac{L_y}{L_x}\,,
\end{equation}
which is the Nyquist result for the current fluctuations (see discussion below).
As seen from Fig.~\ref{fig-5} (b), in the contrast to the voltage fluctuations at the stabilized current
regime shown in Fig~\ref{fig-5} (a), the current fluctuations in the voltage
stabilized regime are always finite, decrease with growing current/voltage drop
reaching a minima at the pinch-off voltage.

\section{Discussion and conclusions.}

The considered model of quasi-ballistic FET admits analytical solutions for
spatially dependent electron concentration, $N(\xi)$,
drift velocity, $V(\xi)$, and  potential, $\Phi(\xi)$, in implicit forms.
 For example, the coordinate dependence $N(\xi)$  is given by Eq.~(\ref{N-x})
(all are dimensionless quantities scaled according to Eqs.~(\ref{sc-1}) -
(\ref{J-B})).
These dependencies are functions of two dimensionless parameters, the flux (current) $J$
and the factor $B$ of Eq.~(\ref{J-B}). The real voltage and current density  can be
calculated by using the scaling parameters $u_{sc}$ of Eq.~(\ref{sc-2})
and $j_{sc}$ of of Eq.~(\ref{J-B}). The current-voltage characteristic of FET,
$J(U)$, is given by Eq.~(\ref{I-V-2}) with the only parameter $\cal B$, which has the
meaning of the dimensionless resistance in the linear operation regime
(see Eq.~(\ref{I-V-3})). Using
notations (\ref{sc-1}),  (\ref{sc-2}) and writing the {\it total current}
in the channel as $- e j_0 L_y$, we found that
$$
 {\cal B}= R \, \frac{e^2 n_s L_y}{m} \sqrt{\frac{m\,\kappa}{4 \pi e^2 h n_s}}
$$
with
\begin{equation} \label{R}
R = \frac{m L_x}{e^2 \tau n_s L_y}
\end{equation}
being the resistance of the conductive channel with carriers characterized by the
scattering time $\tau$ and the mobility $\mu = e \tau/m$.  The applied model predicts
steady-state solutions
only for a finite interval of the current ($J \leq J_c (B))$ and voltage
 ($U \leq U_c$). The value $J_c$ is determined by
Eq.~(\ref{Jc}), while $U_c = J_c^{2/3}$. In Fig.~\ref{fig-2} the filled area
 of the $\{J,B\}$-plane presents allowed currents for this model.
  At $J > J_c$, when the model is not  applicable, a current saturation
regime should  occur. Detailed discussion of the quasi-ballistic FET for $J \geq J_c$ is presented
in paper~\cite{Shur-2}. Here we only recall that at $J \rightarrow J_c$ electron concentration and velocity
remain finite at a finite relaxation time $\tau$.
 Thus, no real pinch effect is realized.
However, approaching $J_c$  the lateral electric field increases infinitely.

In the model, besides the parameter ${\cal B}$ two scaling
parameters $u_{sc}$ and $j_{sc}$ are important.
The latter parameters are determined by the electron characteristics, charge, mass,
concentration, and dielectric surrounding and distance between the channel and the gate, $h$.
Evidently, these parameters reflect
the essentiality of electron-metal gate interaction.
For  numerical estimates, we consider the FET with
the GaAs conductive channel at $77\,K$ and $300\,K$. We set
$m= 0.063\,m_0\,,\kappa =10.9\,,
n_s =10^{12}\,cm^{-2}$ and $h = 5 \times 10^{-7}\,cm$. Then, the scaling parameters
for voltage and current density are $u_{sc} = 83\,mV$ and $e j_{sc} =7.3\,A/cm$.
For $77\,K$ at the mobility $\mu = 3 \times 10^5\,cm^2/V s$,
we find $\tau \approx 10^{-11}\,s$ and $B=1$ at $L_x \approx 3.4\,\mu$.
Thus, numerical results presented for $ {\cal B}= 0.5$ and ${\cal B}=2$ correspond
to the channel lengths $1.7\,\mu m$ and $6.8\,\mu m$ for the low temperature.
For $300\,K$ and $\mu = 800\,cm^2/V s$,
 we find $\tau \approx 3 \times 10^{-13}\,s$ and
${\cal B}=1$ at $L_x= 0.1\,\mu m$. Correspondingly, the numerical results for
${\cal B}=0.5$ and $2$ are for $L_x=0.05\,\mu m$ and $0.2\,\mu m$.

The found steady-state solutions facilitate analysis of low-frequency electron
fluctuations, which originate from random electron scattering in the FET channel.
This analysis was performed applying the Langevin approach
based on the linearized dynamic equation supplemented
by the random force related to electron scattering
(see Eq.~(\ref{L-1})).  Using the microscopic Langevin forces for
bulk electrons characterized by a scattering time, $\tau$, we derived the
random force correlator for hydrodynamic equation of the electron
flux in narrow active layer of FET (see Eq.~(\ref{A-4})). At low frequency
fluctuations ($\omega \tau \ll 1$),
the corresponding Langevin equation (\ref{V-omega})
was solved. Its general solution for spatially dependent fluctuations in
the channel can be calculated analytically using the steady-state
electron distribution $N(\xi)$.
Particular results of the fluctuation analysis are dependent on properties of the
 low-frequency {\it ac} circuit with the FET.
Two cases of such circuits are considered: a circuit with  stabilized
 current (A) and a circuit with stabilized voltage on the FET (B).

For the circuit A, the
analytical expression for the spectral density of low frequency electron velocity
fluctuations is given by Eq.~(\ref{S_V2}). From this
expression it follows that the velocity fluctuations increase along the channel and
reach maximal values at the drain side of the FET. They also
increase with the current through the device (see Fig.~\ref{fig3B} (a)).
 At the current approaching to the critical value, $J \rightarrow J_c$,
the velocity fluctuations grow infinitely. Similar behavior is observed
for spatial distributions of fluctuations of the electron concentration (i.e.,
charge fluctuations in the channel) and the local voltage (see Eq.~(\ref{phi-1}) and
Fig.~\ref{fig3B} (b)).  As for the fluctuations of the voltage drop on the device,
at low current/volatage the spectral density of Eq.~(\ref{phi-2})
coincides with the Nyquist formula. Indeed,
$S^A_{u_{\omega}} =4 D  m^2 L_x / e^2 n_s \tau^2 L_y = 4 k T R$
with $R$ being the low voltage resistance of the active channel (see Eq.~(\ref{R})).
As the current increases, the voltage fluctuations also increase and become
infinitely large at approaching the critical current value, $J \rightarrow J_c$
(see Fig~\ref{fig-5} (a)).

For the circuit B, the spectral density of current fluctuations is given by
 expression (\ref{s-j}) with $N(1)$ determined from Eq.~(\ref{N-x})
 at $\xi=1$ and Eq.~(\ref{I-V-2}). At low voltage Eqs.~(\ref{S-J-0}) and
 (\ref{s-j-2}) recover the Nyquist result for the current fluctuations:
 $s^B_{j}= 4 k T/R$. As the voltage increases intensity of the current fluctuations
 decreases reaching minimal value at $J \rightarrow J_c$ (see Figs.~\ref{fig-5} (b)).
 Though for the circuit B the total voltage drop does not fluctuate,
 the local concentration (charge) and  the potential do fluctuate.
 Spatial distributions  of these fluctuations are not monotonous
 with maximal intensity in a middle of the channel, as shown in
 Fig.~\ref{fig3B} (c).  Note, spatial distributions of charges and electrostatic
potentials and their low-frequency
fluctuations on nanoscale can be studied using different
scanning microscopic methods for specially designed FET with access to surface
of the active channel. For example, nanometer scale
imaging of the surface potential can be received
with the Kelvin probe force microscopy\cite{Kelvin}.
Such a study can provide additional information on transport processes
within the device.

Comparing results for two values of the parameter ${\cal B}$ presented in
Figs.~(\ref{fig-3}) - (\ref{fig-5}) and taking into account the Nyquist
quantities (\ref{phi-1}), (\ref{s-j-3}) relevant to these values,
we can conclude, that at a given current the device with
{\it larger ballisticity} of the channel (${\cal B}=0.5$) demonstrates {\it
larger low-frequency voltage noises} for the circuit A and {\it smaller
current noises} for the circuit B in comparison to more dissipative
channel of ${\cal B} =2$.

 It is pertinent to note that often the Nyquist formulas are used to estimate
 the current or voltage
 noises under nonlinear operation regimes with
substitution of the low-voltage/current resistance, $R$, by the differential
resistance $R_D = d u /d (- e j)$.   For the FET model under
 consideration, the dimensionless differential resistance is given by
 Eq.~(\ref{d-resist}). In  Figs~\ref{fig-5} (a), (b)
 we presented the results of such estimates. It can be seen that the noise
 characteristics obtained in this paper differ significantly from
mentioned above estimates. Moreover, the correct analysis predicts higher fluctuation
intensities for both $A$ and $B$ circuits.

Presented study of noises generated by internal random scattering processes
under nonlinear electron transport replenishes the list of
reported analytical studies of electron fluctuations in device structures with nonuniform
distributions of carriers and fields induced by currents; examples include: noises in
$p-n$-junctions\cite{example-1},   $n^+-n$-junctions~\cite{example-2,example-3,example-4},
excess noise in nonuniform conductive channels~\cite{example-9},
shot noise in injection diodes~\cite{example-5,example-6,example-7,example-8},
hot-electron and intervalley size  effects for fluctuations~\cite{example-10,example-11}, etc.

The analysis makes better comprehension of physics of electron transport and
fluctuations in quasi-ballistic FETs, particularly, revealing the effect of
electron correlations under the metal gate on electrical fluctuations.
The results improve theoretical description of quasi-ballistic FETs,
which is essential from viewpoints of the device simulation and design.

\section*{Author Declarations}

Conflict of Interest
The authors have no conflicts to disclose.

\section*{Author Contributions}

Mykola Yelisieiev: Investigation (equal); Writing - review and editing (equal).
Vyacheslav A. Kochelap: Conceptualization (lead); Supervision (lead); Investigation (equal);  Writing - original draft;

\section*{DATA AVAILABILITY}
The data that support the findings of this study are available
from the corresponding author upon reasonable request.

\appendix
\section{Correlator of random force $f(x,t)$}

The Langevin approach and the above introduced random force, $f(x,t)$, are valid under the
following conditions. First, the ensemble average of a variable ($v_1$ in our case)
over a small time interval $\Delta t$  can be described by the same {\it macroscopic}
dynamic equation (1). Second, this time interval has to be much greater than correlation
time of true {\it microscopic} Langevin source (the force $f({\bf r},t)$ in our case).
The same is assumed for coordinate dependence of the fluctuation source.
 These conditions
allow to set that the averaged over ensemble correlator
of the random  force in the dynamic equation is proportional to $\delta({\bf r}-{\bf r'})\,\delta(t-t')$.
The general relationship for a random force is~\cite{Kogan}:
\begin{equation} \label{A-1}
\overline{f_i({\bf r},t)f_j({\bf r}',t')} = g_{ij}({\bf r}, t)\, \delta({\bf r}-{\bf r}')\, \delta(t-t')\,,
\end{equation}
where $f_i({\bf r},t)$ with $i,j=x,y,z,$ are projections of this vector force,
functions $g_{ij}({\bf r}, t)$ are determined by the local parameters
(distribution over velocities, carrier densities, etc).

The latter general equation can be applied to derive the correlator presented in
Eq.~(\ref{corr-2}). To find the function $g(x)$ used in the equation for the fluctuations over
the steady-state, let us take advantage of the well known expression for the correlator  of flux
fluctuations, $\delta {\cal J}_{3D,i}({\bf r},t)$, for  three-dimensional (3D) electrons provided in~\cite{Kogan}:
\begin{equation} \label{3D-J}
 \overline{\delta {\cal J}_{3D,i}({\bf r},t) \,\delta {\cal J}_{3D,j}({\bf r}',t')} =
 2 D n_{3D}({\bf r})\,\delta_{ij}\, \delta({\bf r}-{\bf r}')\,  \delta(t-t')\,.
\end{equation}
Here  $D$ is a diffusion coefficients, $n_{3D}({\bf r})$ is the electron density.
For nondegenerate electrons one can set
$D=\tau_p \nu_T^2/3$ with $\tau_p (\leq \tau)$ and $\nu_T$ being the collision
time and the thermal velocity of the electrons, respectively.
The density of the electron flux through the cross-section
of the conductive channel, $j$, (see Eq.~(2)) can be obtained by integration of
the bulk flux density, ${\cal J}_i =n_{3D} \nu_i$, over $z$ and averaging over $y$:
$$j = j_x = \frac{1}{L_y} \int dz \int_0^{L_{y}}dy \,  {\cal J}_{3D,x}\,, $$
(see the device geometry in Fig.~\ref{fig-1}).
Applying similar procedure for Eq.~(\ref{3D-J})  we easily obtain
the correlator of the local fluctuations of $j_x$:
\begin{equation} \label{2D-j}
 \overline{\delta j_x (x,t) \,\delta j_x (x',t')} = 2 \frac{D n (x)}{L_y}\, \delta(x-x')\,  \delta(t-t')\,
\end{equation}
with $n$ being the area density:
$$ n (x) = \frac{1}{L_y} \int dz \int_0^{L_{y}}dy  \,n_{3D}({\bf r})\,.$$
Here we assume, that the bulk electron density $n_{3D}$ does not depend on $y,\,z$.
To link the correlator of Eq.~(\ref{3D-J}) and Eq.~(\ref{2D-j}) we again remind that the correlator of
microscopic Langevin sources are determined by the local parameters, $n$ and $D$.
This allows us to apply Eq.~(\ref{L-1}) for the uniform case and to obtain
$\delta j_{x}^{(un)} (x,t) = \frac{\tau n}{m} f (x, t)$ in the limit $\omega \rightarrow 0$.
Now using the correlator of Eq.~(\ref{2D-j}) we find:
\begin{equation} \label{A-4}
\overline{f(x,t) f(x',t')} =
 \frac{2 D\, m^2}{\tau^2 \,n(x)\,L_y}\,\delta(x-x')\,\delta(t-t')\,,
\end{equation}
and
\begin{equation}
g(x) = \frac{2 D\, m^2}{\tau^2\,n(x)\,L_y}\,.
\end{equation}

Finally, we remind the definition of the spectral density of fluctuations of certain physical value $X(t)$:
if correlator of the fluctuations $\delta X(\omega)$ is~\cite{Landau-St-M}
\begin{equation} \label{W-K-1}
2 \pi\,\overline{\delta X(\omega) \delta X(\omega')} =
\delta(\omega+\omega') \Xi (\omega)\,,
\end{equation}
then according to the Wiener Khintchine theorem, the spectral density of the fluctuations equals
\begin{equation} \label{W-K-2}
S_{X_{\omega}} = 2 \Xi (\omega)\,.
\end{equation}

\end{document}